\documentclass[superscriptaddress,showpacs,aps,twocolumn,amsmath,amssymb,apl]{revtex4}

\usepackage{graphicx}
\usepackage{dcolumn}
\usepackage{bm}
\usepackage{mathtext}
\usepackage{amsmath}
\usepackage{xcolor}

\begin{document}

\title{Ultrafast non-thermal laser excitation of gigahertz longitudinal and shear acoustic waves in spin-crossover molecular crystals [$Fe(PM-AzA)_{2}$$(NCS)_{2}$]}

\author{T. Parpiiev}
\affiliation{Institut Mol\'ecules et Mat\'eriaux du Mans, UMR CNRS 6283, Universit\'e du Maine, 72085 Le Mans, France}
\author{M. Servol}
\affiliation{Institut de Physique de Rennes, UMR CNRS 6251, Universit\'e de Rennes 1, 35042 Rennes, France}
\author{M. Lorenc}
\email{maciej.lorenc@univ-rennes1.fr}
\affiliation{Institut de Physique de Rennes, UMR CNRS 6251, Universit\'e de Rennes 1, 35042 Rennes, France}
\author{I. Chaban}
\affiliation{Institut Mol\'ecules et Mat\'eriaux du Mans, UMR CNRS 6283, 
Universit\'e du Maine, 72085 Le Mans, France}
\author{R. Lefort}
\affiliation{Institut de Physique de Rennes, UMR CNRS 6251, Universit\'e de Rennes 1, 35042 Rennes, France}
\author{E. Collet}
\affiliation{Institut de Physique de Rennes, UMR CNRS 6251, Universit\'e de Rennes 1, 35042 Rennes, France}
\author{H. Cailleau}
\affiliation{Institut de Physique de Rennes, UMR CNRS 6251, Universit\'e de Rennes 1, 35042 Rennes, France}
\author{P. Ruello}
\affiliation{Institut Mol\'ecules et Mat\'eriaux du Mans, UMR CNRS 6283, 
Universit\'e du Maine, 72085 Le Mans, France}
\author{N. Daro}
\affiliation{Institut de Chimie de la Mati\`ere Condens\'ee de Bordeaux, UPR CNRS 9048, Universit\'e de Bordeaux, 33608 Pessac, France}
\author{G. Chastanet}
\affiliation{Institut de Chimie de la Mati\`ere Condens\'ee de Bordeaux, UPR CNRS 9048, Universit\'e de Bordeaux, 33608 Pessac, France}
\author{T. Pezeril}
\email{thomas.pezeril@univ-lemans.fr}
\affiliation{Institut Mol\'ecules et Mat\'eriaux du Mans, UMR CNRS 6283, 
Universit\'e du Maine, 72085 Le Mans, France}

\begin{abstract}

We report GHz longitudinal as well as shear acoustic phonons photoexcitation and photodetection using femtosecond laser pulses in a spin-crossover molecular crystal. From our experimental observation of time domain Brillouin scattering triggered by the photoexcitation of acoustic waves across the low-spin (LS) to high-spin (HS) thermal crossover, we reveal a link between molecular spin state and photoexcitation of coherent GHz acoustic phonons. In particular, we experimentally evidence a non-thermal pathway for the laser excitation of GHz phonons. We also provide experimental insights to the optical and mechanical parameters evolving across the LS/HS spin crossover temperature range. 

\end{abstract}


\maketitle 

Understanding how ultrafast photoinduced molecular switching in crystals couples to the lattice in optical materials is one of the key challenges in the fields of ultrafast photo-induced phase transitions or transformations and ultrafast acoustics. Systems like spin-crossover compounds exhibit, with temperature or pressure, change of the molecular spin state in the central d$^4$-d$^7$ transition metals of the complex. They are promising candidates for diverse applications including miniature temperature sensors, displays, data storage and photonic devices \cite{Letard2004, Bousseksou}. Moreover, they can be fabricated in a variety of forms (bulk, powders, matrices, films) and down sizable to nanoparticles \cite{Bousseksou}. The possibility to trigger their properties by light \cite{Hauser99}, especially in ultrafast fashion offers further prospects in future applications as optically controlled switching devices \cite{Bousseksou, Letardpatent}. While such compounds have already been widely studied \cite{Guionneau99, Goujon, Marino2013, Jiang}, there is a shortage of information about the interplay between the change of molecular spin state, the change of the unit cell volume (5\% change) and the covalent bonding (10\% change). From this standpoint the study of crystal deformations that trigger excitation of coherent acoustic phonons clearly deserves further experimental investigations. Until recently \cite{Bertoni2016}, in the field of photo-induced phase transitions or transformations in molecular crystals, the role of coherent optical phonons has been long under scrutiny because of the central role of optical mode softening, whereas that of coherent acoustic phonons and lattice deformations has not benefited from the same surge of effort. It is of paramount importance from the fundamental standpoint, as well as for the control of non-volatile information and energy storage, to further explore the pathways whereby acoustic phonons lead to non-volatile photo-induced states.

\begin{figure}[t!]  
\centerline{\includegraphics[width=9cm]{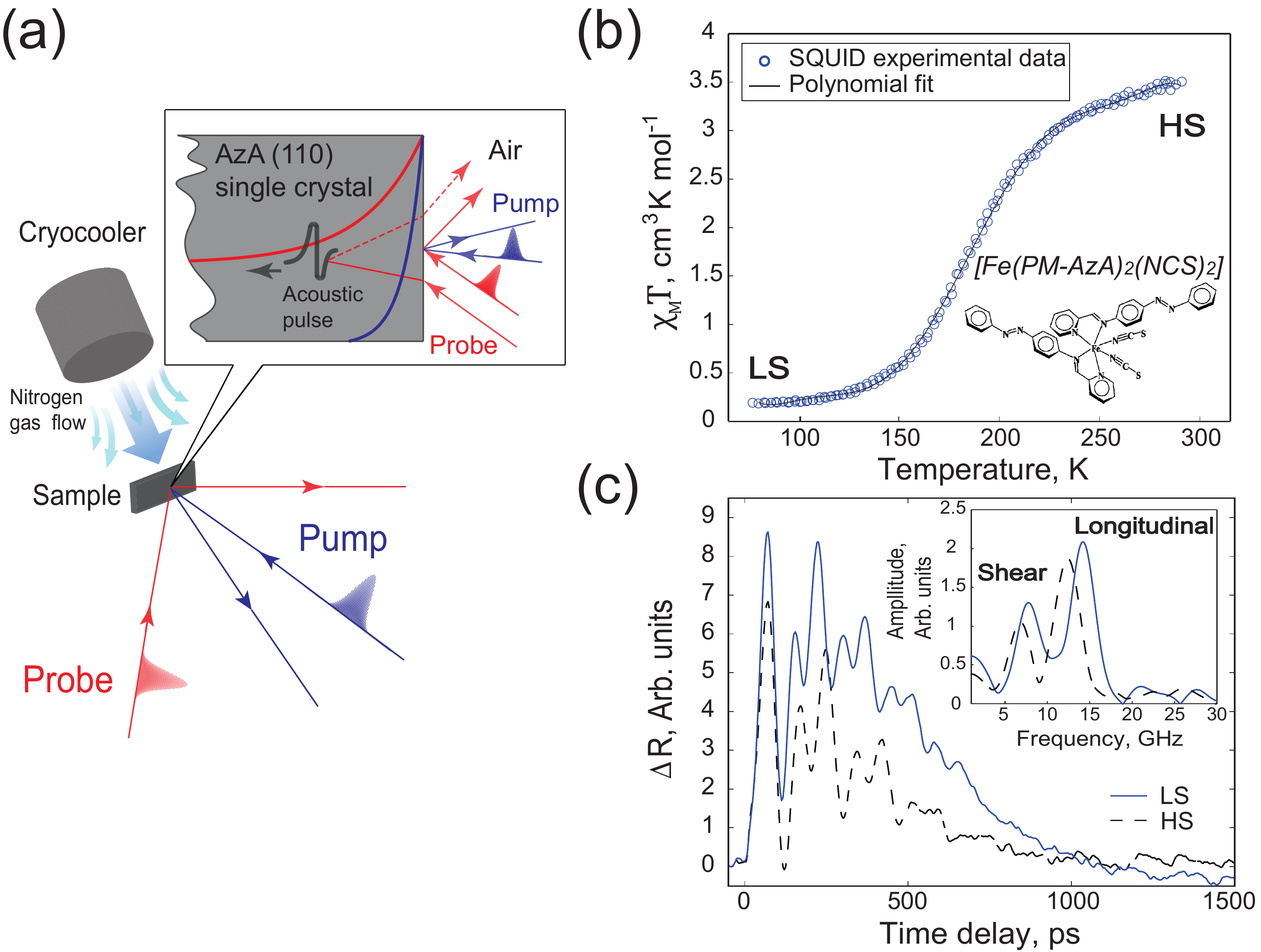}}  
\caption{\label{opto}(Color online) (a) Sketch of the experimental setup. A laser pump pulse photoexcites the molecular crystal sample under continuous nitrogen flow for temperature control. The propagation of the photoexcited acoustic pulse is detected by a time-delayed optical probe pulse vertically polarized. (b) Schematic representation of the spin-crossover compound \cite{Guionneau99} and its magnetic susceptibility recorded across the smooth thermal spin-crossover temperature T$_{1/2}$. (c) Transient reflectivity signals recorded in the LS and HS states. The frequency spectrum of the LS and HS Brillouin oscillations shown in inset gather crucial information on the photoexcitation of both longitudinal and shear acoustic phonons.}
\end{figure}

In the following, we describe our experimental results dealing with the photoexcitation of coherent acoustic phonons in a spin-crossover crystal, performed at different temperatures around the spin crossover temperature T$_{1/2}$. The possibility of exciting with light pulses coherent acoustic phonons in a variety of materials is well known \cite{Thomsen86, Letard99}, however unlike the longitudinal acoustic phonons, the shear acoustic phonons are difficult to photo-excite \cite{Pezeril2007,Pezeril2016}. Our results highlight peculiar and efficient mechanisms for the photoexcitation of shear acoustic phonons and shed light on the interplay between molecular and elastic parameters in a spin-crossover compound.\\ 


In the present experiments sketched in Fig.~1(a), spin-crossover molecular crystals [$Fe(PM-AzA)_{2}(NCS)_{2}$], schematically represented in the inset of Fig.~1(b) were investigated. These molecular crystals belong to the monoclinic space group $P2_{1/c}$ with one molecule as asymmetric unit. In agreement with \cite{Guionneau99, Marino2013}, the smooth thermal spin-crossover of these crystals with a spin crossover around T$_{1/2}\sim$~180~K can be monitored experimentally from the measurement of the magnetic susceptibility $\chi_{_M}$ and the product $\chi_{_M}$T, as indicated on Fig.~1(b). The experiments were performed on a parallelepipedic 4$\times$1$\times$1 mm$^3$\ single crystal, with smooth, black color surfaces \cite{Marino2013}. Due to large optical absorption by the crystal, front side pump-probe transient reflectivity measurements were performed, see Fig.~1(a). The pump and probe beams originate from a femtosecond Ti-Sapphire Coherent RegA 9000 regenerative amplifier operating at central wavelength of 800~nm and delivering 160~fs pulses at a repetition rate of 250~kHz. The 400~nm pump pulses of about 40~nJ energy per pulse were focused on the (110) surface of the crystal with a gaussian spatial profile of FWHM $\sim$ 100~$\mu$m. The time delayed 800~nm probe of tenfold weaker energy per pulse, vertically polarized, was tightly focused at 30$^\circ$ oblique incidence on the (110) surface normal to the crystal and spatially overlapped with the pump spot. The reflected probe beam was directed to a photodiode coupled to a lock-in amplifier operating at 50~kHz of the pump laser modulation frequency, to measure transient differential reflectivity $\Delta$R(t) as a function of time delay between pump and probe beams. Upon transient absorption of the 400~nm pump pulse over the optical skin depth of the crystal, the light energy is partially converted into mechanical energy that drives the excitation of acoustic pulses propagating away from the free surface. In the present situation, the crystal is about five times less absorptive at 800~nm than at 400~nm, see Fig.~S1 in supplementary material, such that it can be considered as semi-transparent at 800~nm and opaque at 400~nm. As a consequence, the pump light is locally absorbed at the free surface where it launches a propagating acoustic strain that back scatters the probe light from within the semi-transparent medium and leads to the occurrence of time domain Brillouin scattering oscillations, see Fig.~1(c). As in any Brillouin scattering process, the frequency $\nu$ of these oscillations is related to the ultrasound velocity $v$ of the crystal, to the probe wavelength $\lambda$, to the refractive index $n$ of the medium, and to the back-scattering angle $\theta$ through
\begin{equation}\label{Brillouin}
\nu = 2 \ n \ v \cos \theta / \lambda.
  \end{equation}
The light activation of both longitudinal and shear acoustic polarizations, of different ultrasonic speeds $v$, leads to two distinct Brillouin frequencies. Fig.~1(c) shows an example of such time domain Brillouin scattering light modulation where unambiguous periodic features at about 6~GHz and 12~GHz, see inset, are evidenced right after pump excitation at zero time delay. The shear acoustic nature of the 6~GHz frequency has been further experimentally confirmed from depolarized Brillouin scattering measurements, see Fig.~S2 in the supplementary material, which enhances the optical detection of shear acoustic waves \cite{Pezeril2016}.

\begin{figure}[t!]  
\centerline{\includegraphics[width=9cm]{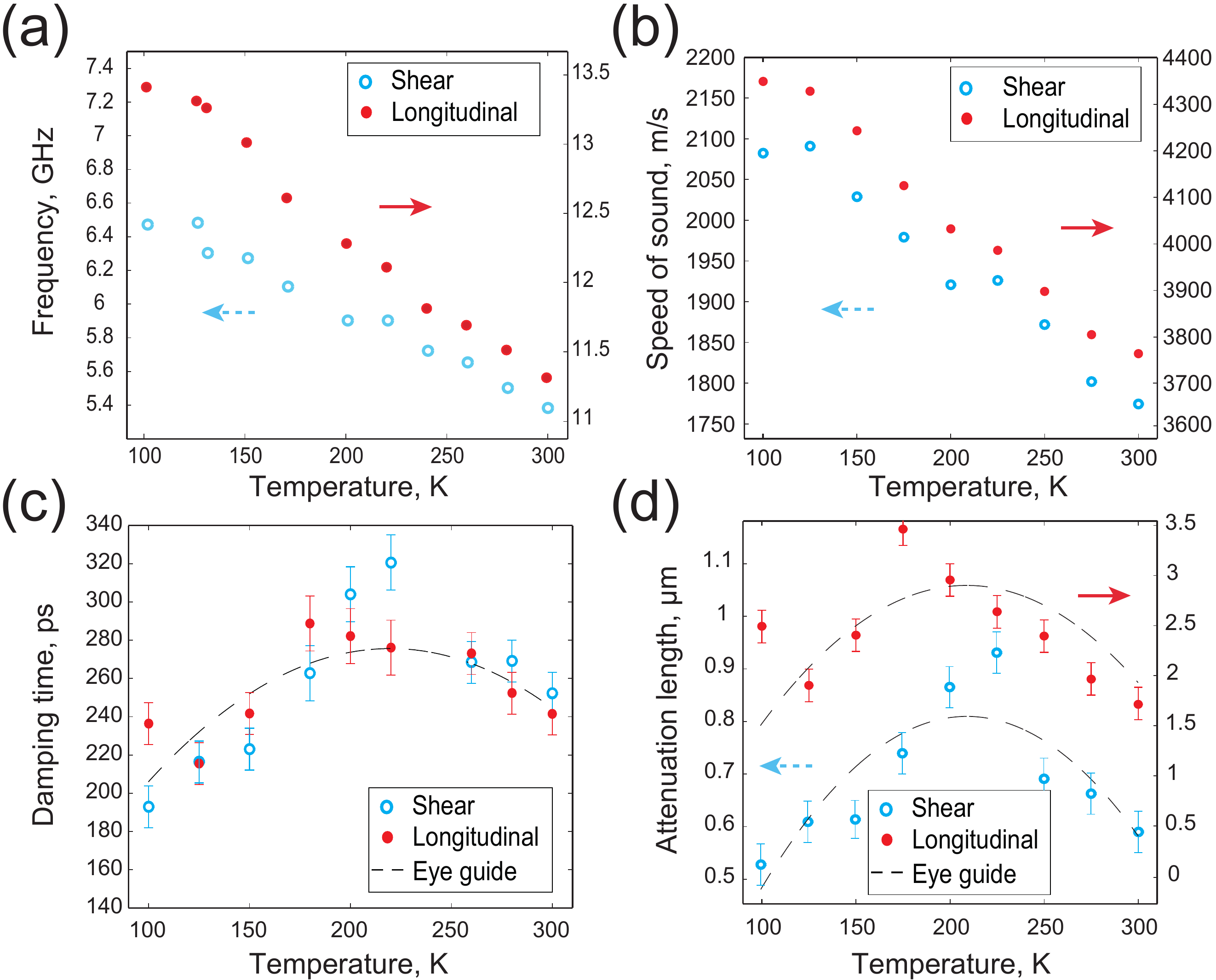}} 
\caption{\label{Fig2}(Color online) From the time domain Brillouin scattering experimental data obtained for vertically polarized pump and probe beams at different temperatures, we have extracted the frequency (a) and the damping coefficient (c) of the measured Brilllouin oscillations. In a second step, the temperature evolution of the index of refraction in the supplementary material led to the calculated acoustic speed (b) and acoustic attenuation length (d) of the longitudinal and shear acoustic modes across T$_{1/2}$. The eye guides are 2nd order polynomial fit of the extracted coefficients and the error bars are estimated from our experimental uncertainties.}  
\end{figure}

The experiments have been conducted at different temperatures ranging from 100~K, where almost 100\% of molecules are in the low spin state, up to 300~K, where almost 100\% of the molecules are in the high-spin state. Temperature steps of 10~K were performed with a continuous nitrogen flow. For each recorded time domain Brillouin scattering signal in the 100 - 300~K temperature range, we have numerically fitted the experimental data with a damped sinusoidal function in the form $\sim A \exp(-t/\tau)\sin(2\pi\nu t+\phi)$, to retrieve the frequency $\nu$, the damping time $\tau$, the amplitude $A$, and the phase $\phi$ of each longitudinal and shear acoustic mode. Based on the results from ellipsometry available in the supplementary material, which led to the observation of a slight variation of only a few percent of the optical refractive index of the crystal at different temperatures, the huge 15\% change in Brillouin frequency in Fig.~2(a) is mainly due to a pronounced change in both longitudinal and shear acoustic speed with temperature. According to Eq.~(\ref{Brillouin}) and the measured real part n$_{\perp}$ of the index of refraction at the probe wavelength for vertically polarized light, see supplementary material, we have computed the change in Brillouin frequency to extract the variation in longitudinal and shear acoustic speed with temperature. The results displayed in Fig.~2(b) confirm the substantial change in acoustic speed, in the range of 10\%, further emphasizing the giant change in mechanical properties of the material across the spin crossover temperature. This intrinsic softening of the material at increasing temperature around T$_{1/2}$ is coupled to a substantial isostructural modification of the lattice parameters, manifested in the change of the unit cell volume by as much as $\sim$3 \% \cite{Guionneau99,Klinduhov2010}. 

As a comparison with highly magnetostrictives ferromagnetic compounds such as Terfenol \cite{Kovalenko13} which is the foremost highest magnetostrictive alloy with a change of the unit cell volume by an amount of 0.1 \% upon modification of the magnetization vector, the change in [$Fe(PM-AzA)_{2}(NCS)_{2}$] lattice parameters in the order of few \% reveal the gigantic spin state-lattice coupling in these molecular materials. The pronounced molecular spin state-lattice coupling in such compounds is bound to influence the coherent acoustic phonons generation.

The evolutions of the damping time coefficients with temperature displayed in Fig.~2(c) are linked to the imaginary part of the refractive index k$_\perp$ at the probe wavelength and to the intrinsic acoustic attenuation. From the measured k$_\perp$ at the probe wavelength, we have processed the measured damping of the Brillouin oscillations $\tau$, in order to extract the acoustic attenuation length $\Gamma$ straightforwardly from $1/\Gamma$~=~$1/v\,\tau$~-$1/\xi$ where $v$ is the longitudinal or shear acoustic speed and $\xi=\lambda/4\pi k_\perp$ is the optical penetration depth. Note that the deconvolution of the acoustic damping in the measured signals is not as straightforward in classical Brillouin spectroscopy which sometimes requires complicated analyses of the Brillouin linewidth of the peaks in the frequency domain \cite{Sandercock}. The result of the calculation of the acoustic attenuation length for both acoustic modes, longitudinal and shear, is displayed in Fig.~2(d). The remarkable maximum attenuation length measured in Fig.~2(d) highlights a decrease of the acoustic attenuation across the spin-phase crossover transition. In the present case, the phenomenon is reversed from the well-known structural $\alpha$-relaxation which has been evidenced in glass forming liquids across the T$_g$ glass transition temperature \cite{Yang}. The acoustic wave propagates on longer distances at the spin-crossover temperature which indicates that the structural modification of the lattice and the statistical growth or disappearance of the LS/HS states do not perturb the acoustic phonons propagation, on the contrary, the acoustic phonons propagation is facilitated during the spin crossover transition.

\begin{figure}[t!]  
\centerline{\includegraphics[width=9cm]{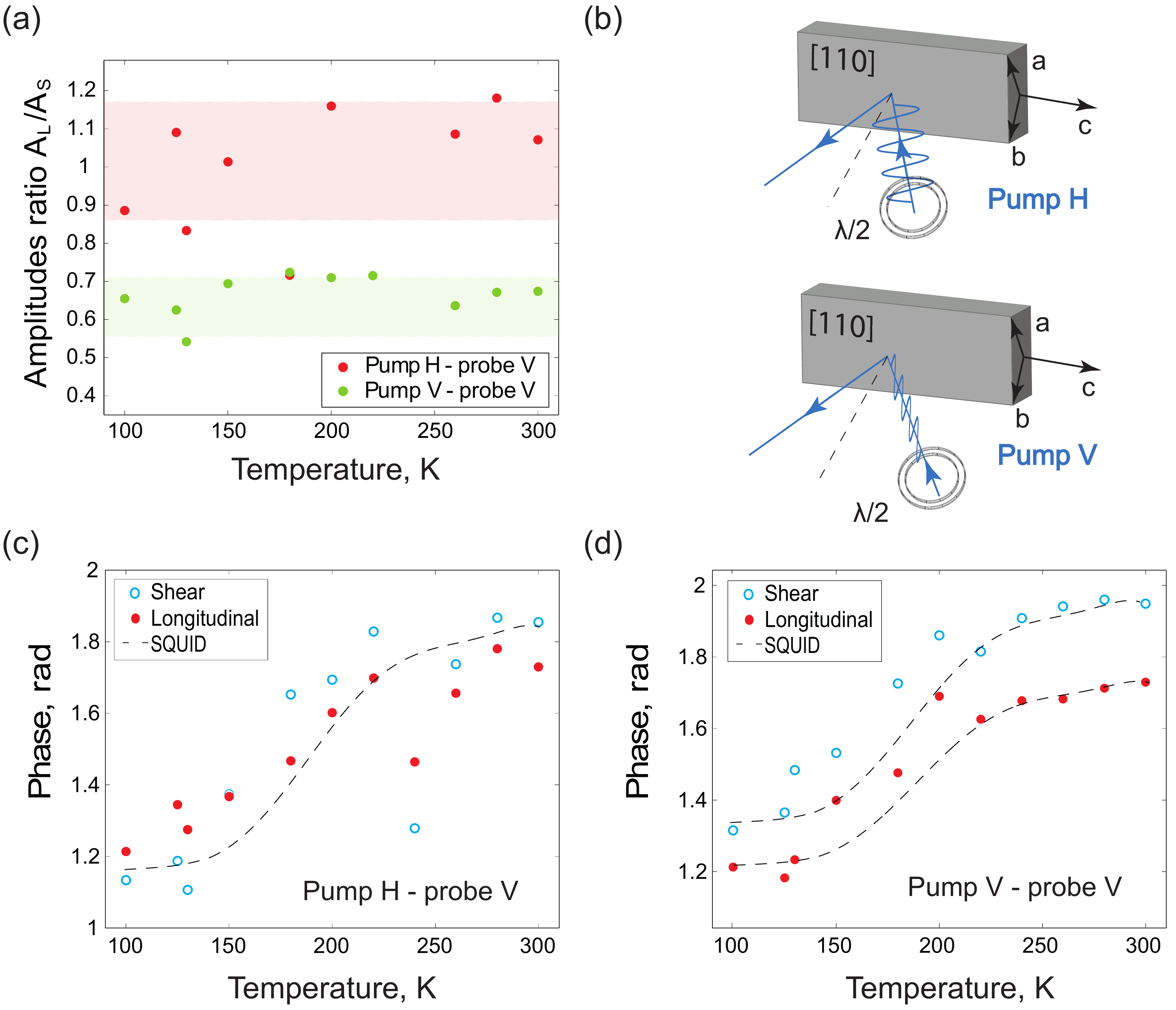}} 
\caption{\label{Fig3}(Color online) (a) Ratio between the longitudinal and shear Brillouin amplitudes for different pump polarizations. Red dots correspond to horizontally polarized pump, green to vertically polarized pump, as sketched in (b). Extracted time domain Brillouin phase for horizontally (c) and vertically (d) polarized pump beams of both longitudinal and shear acoustic modes in the compound at different crystal temperatures. The SQUID curve matches the Brillouin phase evolution across T$_{1/2}$.}
\end{figure}

We can invoke several main mechanisms for the laser-driven lattice motion in the present spin-crossover material: the thermoelastic mechanism which is linked to the transient thermal dilation of the lattice following the temperature rise due to laser absorption, and two non-thermal mechanisms, namely the molecular spin state-lattice coupling mechanism \cite{Enachescu} and the deformation potential mechanism \cite{Pezeril2016}. The temperature evolution of the Brillouin amplitudes and Brillouin phases at two different pump polarizations (horizontal (H) and vertical (V) polarizations) displayed in Fig.~3 gather crucial information on the photoacoustic excitation process. The fact that the optical index of refraction does not vary significantly with temperature warrants that the measured Brillouin amplitude and phase are mainly sensitive to the excitation mechanisms and not to the detection process through a modification of the acousto-optic coefficients with temperature. Since the measurement of the Brillouin amplitude can suffer from experimental artifacts, such as beam pointing stability during sample heating or cooling, we have chosen to further process the longitudinal $A_L$ and shear $A_S$ Brillouin amplitudes data by taking the ratio of both amplitudes, defined as $A_L/A_S$, not biased by optical artifacts. This simple procedure highlights a discrepancy shown in Fig.~3(a) that cannot be assigned to the thermoelastic mechanism. In fact, as indicated by the difference in optical absorption coefficients $k_\perp$ and $k_\parallel$ at the pump wavelength, see supplementary material, the laser induced temperature rise in such anisotropic crystals depends on the pump polarization. However, taking the ratio of amplitudes of both modes is a proper way to conveniently remove the temperature rise contribution in the thermoelastic process of acoustic excitation, similar for longitudinal and shear acoustic modes. Therefore the gap between the amplitude ratio in Fig.~3(a) with H or V polarizations is assigned to a non-thermal mechanism, either molecular spin state-lattice coupling or deformation potential mechanism. One possible explanation of the observed amplitude jump would be the anisotropy in deformation potential mechanism. The latter should be considered as tensorial in such crystal with different diagonal and non-diagonal coefficients referring to preferential electronic excitations of the ligands by horizontally or vertically polarized pump pulses, with some similarities with \cite{Rury2016} that demonstrates the wide range of electronic excitations of organic molecules that can drive coherent lattice phonon excitation. In addition to the electron deformation potential mechanism, we cannot neglect the molecular spin state-lattice coupling anisotropy, thus revealing pump laser polarization dependence.

Another striking feature suggesting coexistence of photoacoustic mechanisms is revealed by the Brillouin phase change in Fig.~3(c) and 3(d) with H or V pump polarizations. Once again, if we assume that the slight change with temperature of the optical index of refraction is irrelevant for the interpretation of our experimental observations, the substantial phase jump in the range of 0.6~radians of both longitudinal and shear Brillouin signals across the spin-crossover transition highlights a profound change in the laser-matter mechanism for acoustic phonons excitation. Based on \cite{Vaudel2014}, we can interpret this phase change, which correlates with the magnetic susceptibility of Fig.~1(b), as a change in the acoustic excitation process through the contribution of the spin state-lattice mechanism that vanishes once the compound reaches 100\% HS spin. However, the photoinduced spin state-lattice coupling is maybe not the most efficient at this pump wavelength. Therefore, we cannot rule out the deformation potential mechanism as relevant in the process of laser excitation of coherent acoustic phonons in the present spin crossover material. The observation of phase jump of about 0.2~radians in Fig.~3(d) between longitudinal and shear excitation points to a non-thermal mechanism of acoustic excitation sensitive to the pump polarization, as in Fig.~3(a). As a matter of fact, based on the results presented in Fig.~3, we can conclude that two non-thermal mechanisms are triggered in these molecular crystals, each one having different efficiency (amplitudes) and characteristic times (phases).

In summary, we have performed ultrafast time-domain Brillouin scattering experiments to study non equilibrium dynamics following femtosecond photoexcitation of spin-crossover molecular crystals [$Fe(PM-AzA)_{2}$$(NCS)_{2}$]. We have presented results for coherent GHz acoustic phonons photogeneration and photodetection in a spin-crossover material across the spin-crossover temperature range. Through time domain Brillouin scattering, we evidence non-thermal excitation of acoustic phonons which are of spin state-lattice coupling and/or deformation potential origin. Experimentally revealed on the sub-nanosecond time-scale, remarkable sensitivity of Brillouin frequencies to the spin state of a molecular material, opens advanced perspectives for probing macroscopically relevant processes during a phase transition. We envisage pump-pump-probe experiments, in which molecular spin-state is photoexcited with wavelength tuned pump pulse, while the real-time coupling of thus generated spin-states to the lattice is followed by the second pump through time-resolved Brillouin scattering, such as this work. Furthermore, our results highlight the versatile and efficient generation of ultrashort shear acoustic phonons for future investigations of viscoelastic properties of materials such as liquids \cite{Pezeril2009}, glasses \cite{Klieber2012,Klieber2013}, mixed multiferroics, correlated electron systems and magnetic materials. Ultimately, deeper knowledge of the spin state-elastic coupling in spin-crossover molecular crystals will be crucial for the design of multifunctional molecular devices. 


See supplementary material for the temperature data of the real $n$ and imaginary $k$ refractive index of [$Fe(PM-AzA)_{2}$$(NCS)_{2}$] at 400~nm and 800~nm wavelengths, as for a comparison between transient reflectivity and depolarized Brillouin scattering.


The authors are thankful to Pr. Vitalyi Gusev for beneficial scientific discussions and guidance and to Lionel Guilmeau for technical support. The authors gratefully acknowledge Agence Nationale de la Recherche for financial support under grants ANR-16-CE30-0018, ANR-14-CE26-0008 and ANR-12-BS09-0031-01.

\end{document}